\documentstyle[preprint,aps]{revtex}
\newcommand{\be}{\begin{eqnarray}}
\newcommand{\ee}{\end{eqnarray}}
\newcommand{\ra}{\rightarrow}
\begin{document}
\draft
\preprint{KAIST--CHEP--95/02}
\title{  Strategy for detecting $b$ polarization effects }
\author{
Jae Kwan Kim
and
Yeong Gyun Kim\thanks{ygkim@chep6.kaist.ac.kr},
}
\vspace{.6in}
\address{
Dept. of Physics, KAIST, Taejon 305-701, KOREA
}
\date{\today}
\maketitle
\begin{abstract}
\\
 We suggest a strategy for detecting $b$ polarization effects in $e^+e^-$
annihilation on $Z$ resonance.
Using two types of inclusive leptonic samples
with different
${\Lambda_b}$ fractions in $Z \ra b\bar{b}$ events,
the $b$ polarization effects may be detected
without large uncertainties of $b$ fragmentation and
${\Lambda_b}$ decay model.
\end{abstract}
\pacs{ }

\narrowtext


 The standard model predicts that $b$ quarks are produced with a large
longitudinal polarization in $e^+e^-$ collision on the  $Z$ peak\cite{bpol}.
The measurable $b$ polarization effects are reduced
by the fragmentation
properties of $b$ quarks.
All spin information is lost in the formation of both
pseudoscalar and vector $B$ mesons.
Only $b$-baryons are expected to retain some
spin information\cite{close}.

Various methods which would be used to detect the $b$ polarization effects
have
been suggested by several authors\cite{mele,bonvicini}.
Mele and Altarelli \cite{mele} have suggested that inclusive lepton
energy spectrum in the semileptonic decays of $b$ hadrons could be used
as a measure of $b$ quark polarization. In this
method, a good control of fragmentation effects is crucial and this may be
achieved by calibration on low energy data.
Also one can use the ${\Lambda_b}$ sample of semileptonic events
and compare it
with the $B$-meson sample in order to control fragmentation effects.
In latter case, a possible observed difference
between the ${\Lambda_b}$ and the
inclusive spectra could be blamed on the model dependent features of the
exclusive decay and so
a reliable model of ${\Lambda_b}$ exclusive decay is required.

 In this Brief Report we suggest another possible strategy for detecting $b$
polarization effects, which would be almost free from the uncertainties of
$b$-fragmentation and ${\Lambda_b}$ decay model.
The main point of this strategy is that we study
polarization effects in two types of `inclusive' leptonic samples with
different ${\Lambda_b}$ fractions.

 First, we choose dilepton events in
the $Z \ra b\bar{b}$ events,
in which both $B$ hadrons decay semileptonically.
We can catalog
these events into the ``like-sign events'' in which leptons have same charge
and the ``unlike-sign events'' in which leptons have different charge.
Clearly the existence of the like-sign events is
a signal of $B-\bar{B}$ mixing\cite{mixing}.
In these events, one
of the $B$ hadrons must be a $B_s$ or $B_d$ meson.
However this is not the case in
the unlike-sign events.
So the fractions of ${\Lambda_b}$ would be different  for each
types of events and are given by
\be
F^{like}({\Lambda_b}) =\frac{f_{\Lambda_b}} {2(1-{\chi_B})}
{}~~~~~~~~~~~~~~~ for~~ `like~ sign~~ events',
\ee
\be
F^{unlike}({\Lambda_b}) =
\frac{f_{\Lambda_b}(1-{\chi_B})} {[(1-{\chi_B})^2+{\chi_B}^2]}
{}~~~~~~ for~~ `unlike~ sign~~ events'
\ee
where $f_{\Lambda_b}$ is the branching ratio for
${\Lambda_b}$ production
from  $~b$  quark and
 ${\chi_B}$  is  $~B$  hadron mixing parameter.
 We notice that the ratio
$F^{unlike}({\Lambda_b})/F^{like}({\Lambda_b})$ is independent of
$f_{\Lambda_b}$ and
would be almost two with small value of ${\chi_B}$.
 The mixing parameter ${\chi_B}$
gives the probability that a hadron containing $b$ quark oscillates
into a hadron
containing a $\bar{b}$ quark at the time of its decay
and it has been measured at
$\sqrt{s}=M_Z$\cite{mixing}.
If we take the value $f_{\Lambda_b}$ = 0.1 and ${\chi_B}$ = 0.12, then we get
$F^{like}({\Lambda_b})=5.7\%$ , $F^{unlike}({\Lambda_b})=11.2\%$.
 So we have two types of inclusive leptonic samples with
different ${\Lambda_b}$ fractions.

 In the next step, we calculate the ratio $R^{N}_{\l}({\hat P})$
of the $N$-th moments
of the lepton energy spectrum
in the unlike-sign sample over the ones in the like-sign sample,
\be
R ^{N}_{\l}({\hat P}) = \frac{{\langle x^N_{\l} \rangle}({\hat P})}
{{\langle x^N_{\l} \rangle}({0.5 \hat P})}
\ee
where $x_l=2 E_l/ {\sqrt{s}}$, $~E_{\l}$ is lepton energy in lab. frame
, $\sqrt{s}$ is c.m. energy of $e^+e^-$ collision,
and ${\hat P}$ is the effective polarization of
the unlike-sign sample. Here, we assumed
the ratio $F^{like}({\Lambda_b})/F^{unlike}({\Lambda_b})$
is 0.5  and this would be a good approximate value with the measured
value of ${\chi_B}$\cite{mixing}.
Then the effective polarization of
the like-sign sample should be $0.5 \hat P$.
In the collinear approximation,
the observed leptonic spectrum in the laboratory
is the convolution of the polarization-dependent
`naive' leptonic spectrum with
$b$ fragmentation function.
Then,
\be
R ^{N}_{\l}({\hat P}) = \frac{{\langle x^N_{\l} \rangle}({\hat P})}
{{\langle x^N_{\l} \rangle}({0.5 \hat P})}
=\frac{\langle x^N_{b} \rangle {\langle x^N \rangle}({\hat P})}
{\langle x^N_{b}\rangle {\langle x^N \rangle}({0.5 \hat P})}
= \frac{{\langle x^N \rangle}({\hat P})}
{{\langle x^N \rangle}({0.5 \hat P})}
\ee
where $x_b=2 E_b/{\sqrt{s}}$ , $x=E_l/E_b$ and
$E_b$ is $b$ quark energy in lab. frame.
(see ref.\cite{mele} for relevant definitions).
Here, b-fragmentation effects are mostly cancelled and
only the ratio of the moments of the `naive' lepton spectrum
in the unlike-sign sample
over the like-sign sample is left.
The inclusive $b \ra c l {\nu}$ decay can be treated
in the heavy-quark limit as a free-quark decay and there are no nonperturbative
corrections of order ${\Lambda_{QCD}/m_b}$
\cite{chay}. So the above ratio can be predicted reliably.
In fig.1, we present the ratio $R^{N}_{\l}({\hat P})$ for the first three
moments. We used eq. (4) in ref\cite{mele} and
we have taken $m_b$=5 GeV, $m_c$=1.5 GeV.

In the experiment of CERN $e^+ e^-$ collider LEP
, the energy(or momentum) of leptons
is measured with high precision and already a number of
$Z \ra b\bar{b}$ events are avalible. So the above ratio
$R^{N}_{\l}({\hat P})$ could be measured with high precision
and served as the indicator of existence of $b$ polarization.

In conclusion, we have suggested a possible method for detecting $b$
polarization
effects in $e^+e^-$ collision on $Z$ resonance.
We could get two types of
inclusive leptonic samples with different ${\Lambda_b}$ fraction
and calculated
the ratio $R^N_{\l}({\hat P})$ for the first three moments.
This ratio would be almost free from
the uncertainties of $b$ fragmentation and
${\Lambda_b}$ decay model.

\acknowledgements

This work was supported in part by Korea Science and Engineering Foundation
(KOSEF).
\newpage
%

%
%
\newpage
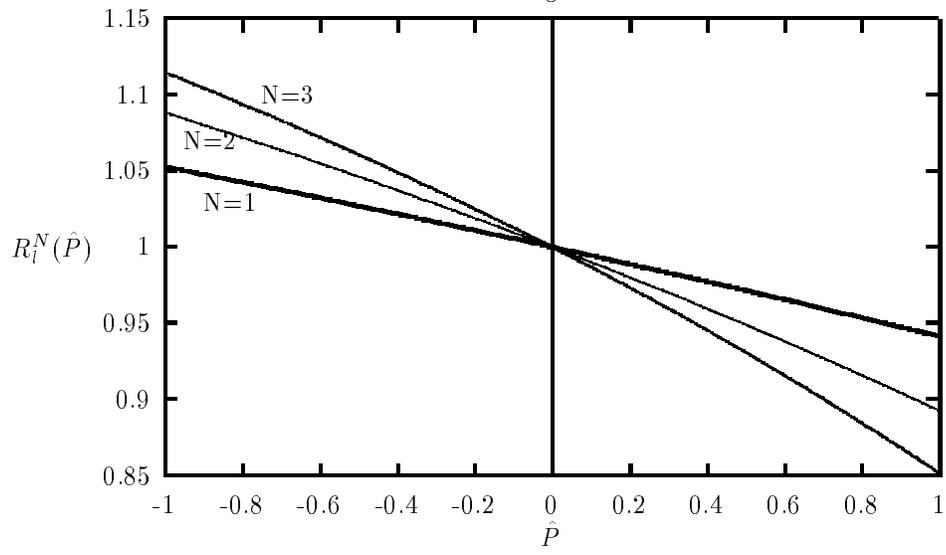
\begin{figure}
\caption{
Prediction for the ratios $R^N_{l}({\hat P})$ with $N$=1, 2, 3 versus
the "effective" b polarization ${\hat P}$.
}
\label{figone}
\end{figure}


\begin{references}
\bibitem{bpol} J.H. Kuhn and P.M. Zerwas, in {\it Heavy Flavors},
eds, A.J. Buras and M. Lindner, (World Scientific, Singapore, 1992), P. 434.
\bibitem{close} F.E.Close, J.Korner, R.J.N.Phillips and D.J.Summers,
J. Phys. {\bf G 18} (1992) 1716.
\bibitem{mele} B.Mele and G.Altarelli, Phys. Lett. {\bf B 299} (1993) 345.
\bibitem{bonvicini} G.Bonvicini and L.Randall, Phys. Rev. Lett. {\bf 73} (1994)
392.
\bibitem{mixing} A. Adam {\it et al.}, L3 Collaboration,
Phys. Lett. {\bf B 335}, 542 (1994).
\bibitem{chay} J. Chay, H. Georgi and B. Grinstein, Phys. Lett. {\bf B 247}
(1990) 399.
\end{references}
\end{document}